\documentclass[11pt,a4paper]{article}
\usepackage{graphicx}
\begin{document}

\title{FINITENESS OF THE HOPPING INDUCED ENERGY CORRECTIONS IN CUPRATES}

\author{GH. ADAM\,$\sp{1,2}$, S. ADAM\,$\sp{1,2}$\\
{\small
$^1$ Horia Hulubei National Institute for Physics and Nuclear Engineering
}\\
{\small
      (IFIN-HH), 407 Atomistilor, Magurele - Bucharest, 077125, Romania
}\\
{\small
      E-mail: adamg@theory.nipne.ro
}\\
\vspace{1mm}
{\small
$^2$ Joint Institute for Nuclear Research, 141980 Dubna, Moscow reg., Russia
}}

\date{}
\maketitle
\begin{center}
(Received \today )
\end{center}

\begin{abstract}
The paper continues the rigorous investigations of the mean field Green
function solution of the effective two-dimensional two-band Hubbard model
[N.M.~Plakida et al., Phys.~Rev.~B, {\bf 51}, 16599 (1995)]
of the superconducting phase transitions in  cuprates, started in
[Gh.~Adam, S.~Adam, J.Phys.A: Math.~Theor., {\bf 40}, 11205 (2007)].
Discussion of the $(\delta, T)$ phase diagram of the model points to
the divergence of the energy spectrum in the limit of vanishing
doping~$\delta$.
Finite energy spectra at all possible doping rates $\delta$ are obtained
provided the hopping part of the effective Hamiltonian is renormalized
with an effective factor pointing to the site-pairs availability for fermion
hopping processes.
\end{abstract}

{\centering\section{INTRODUCTION\label{sec:intro}}}
\smallskip
More than two decades after the discovery of the high critical temperature
superconductivity in cuprates, a theoretical model able to describe
consistently the corresponding phase transition is still missing.
An attractive candidate is the effective two-dimensional two-band Hubbard
model derived by Plakida et al.
\cite{Pl95}
from the more cumbersome three-band $p$-$d$ model
\cite{Emery87,Varma87}
by use of projection techniques based on cell-cluster perturbation theories
\cite{Feiner96a}-\cite{Yushankhai97}
which provide a hierarchy of the various interaction terms.
\par
The present paper is devoted to the scrutiny of the two-band Hubbard model
from the point of view of its consistency with a number of essential features
of the cuprates.
The starting point of this investigation is provided by the rigorous results
reported
\cite{AA07}
for the generalized mean field approximation (GMFA) solution of the
thermodynamic Green function (GF) matrix of the model
\cite{Pl97,Pl03}.
We find that, in order to get consistent solutions everywhere within
a given \emph{cuprate family}, including its reference stoichiometric
structure, the doping rate $\delta$ is to enter explicitly the model
Hamiltonian as a renormalization factor of the Hubbard 1-forms which define
the fermion hopping conduction neighbourhoods of the spin lattice sites.
\par
The paper starts with the discussion of the cuprate features which are
explicitly included into the basic model hypotheses, as well as with other
features which have to be accommodated within the model solution to get a
consistent description of the properties of the cuprates
(section~\ref{sec:essf}).
The Hamiltonian of the model
\cite{Pl95}
is then rewritten in an algebraically equivalent form (section~\ref{sec:H})
which allows easy identification of the various contributions to the
GMFA-GF solution.
The rigorous solution of the GMFA-GF matrix of the model
is rewritten in an alternative form which explicitly points to the energy
spectrum divergence in the limit of vanishingly small doping rates $\delta$
(section~\ref{sec:gmfa}).
A way to secure spectrum finiteness over the cuprate families is discussed
in section~\ref{sec:mhH}.
The paper ends with conclusions in section~\ref{sec:concl}.
\bigskip
\par
{\centering\section{ESSENTIAL FEATURES OF CUPRATES\label{sec:essf}}}
\smallskip

{\centering\subsection{CuO$\sb{2}$ planes embedded in lamellar structures
\label{sec:lams}}}

Irrespective of the specific details of their chemical composition, the
cuprates show \emph{lamellar perovskite structures\/} in the $(a, b)$ plane,
carrying as distinctive elements the \emph{{\rm CuO$\sb{2}$} planes},
separated from each other along the $c$-axis by intermediate planes or other
structures of various compositions. Some of these intermediate structures
have the role to cement the crystal structure, while other
(charge reservoir layers in some cuprates, chain layers in other cuprates
\cite{08Hb07})
provide means for the manipulation of the doping level inside the
CuO$\sb{2}$ planes which are generally assumed to bring the
overwhelming contribution to the superconductivity in cuprates.
\par
Taking into account the weak connection with each other of the CuO$\sb{2}$
planes, the quasi-two-dimensional behaviour of the superconducting
properties of the cuprates is simplified, within the effective two-dimensional
two-band Hubbard model, to a single CuO$\sb{2}$ plane.
Then the effects coming from the weak inter-planar interaction among the
CuO$\sb{2}$ planes are incorporated via effective parameters into the model.
\par
While Geballe and Koster
\cite{08Hb07}
speculate about possible important contributions to the superconducting
pairing coming from structures lying outside the CuO$\sb{2}$ planes, the
correct estimate of their relative magnitude cannot be obtained unless the
quantitative CuO$\sb{2}$ plane contributions are estimated.
\smallskip

{\centering\subsection{Existence of the Fermi surface\label{sec:FS}}}

Fermi surface sheets in well prepared cuprate single crystals have been
undoubtedly evidenced, first, by 2D-ACAR positron spectroscopy at the beginning
of the nineties
\cite{H91}--\cite{A93b}
and then by ARPES and optical methods
\cite{D03}
This result is of paramount importance for the detailed description of the
superconducting phase transition.
The existence of the Fermi surface selects from the energy band structure
contributed by the outer $d$-electron levels  of the Cu ions and the outer
$p$-electron levels of the O ions entering the CuO$\sb{2}$ planes
only those energy bands which lay nearest to the Fermi level of the cuprates.
\smallskip

{\centering\subsection{The charge transfer insulator nature of cuprates
 \label{sec:cti}}}

The identification of the energy bands staying nearest to the Fermi surface
follows from the hierarchy of the interactions resulting in energy level
splittings and hybridizations inside these compounds.
\par
The \emph{crystal field effects\/} coming from the symmetries of the
crystalline structure at the Cu and O sites inside the CuO$\sb{2}$ planes
determine specific splittings of the Cu $3d$ levels and O $2p$ levels
\cite{D03}.
In the hole representation, where the vacuum state is defined by filled
Cu $3d\sp{10}$ and O $2p\sp{6}$ states, the resulting copper one-hole
$d\sb{x\sp{2}-y\sp{2}}$ state and the two oxygen one-hole $p\sb{\sigma}(x)$
and $p\sb{\sigma}(y)$ states inside a given elementary cell belong to a same
irreducible representation of the point group symmetry of the cuprate.
They will therefore \emph{hybridize\/} among themselves resulting in
characteristic energy band structures around the Fermi level.
\par
The three parameters which determine the main features of this energy band
structure (U -- the strength of the electrostatic repulsion among the
electrons, $\Delta$ -- the $p$-$d$ interband splitting, and $W$ -- the
band width) satisfy $U > \Delta > W$, which points to the \emph{charge
transfer insulator\/} nature
\cite{Z85}
of the cuprates.
Two important consequences follow.
\par
First, the hybridization of the copper $d\sb{x\sp{2}-y\sp{2}}$
hole with the oxygen $p\sb{\sigma}(x)$ and $p\sb{\sigma}(y)$ holes results
in two one-hole mixed (predominantly oxygen) $p$-$d$ states.
This interaction inside the elementary cell of the cuprate singles out
the characteristic Zhang-Rice (ZR) singlet
\cite{ZhR88},
such that the two energy bands lying nearest to the Fermi level are the
ZR band and the upper Hubbard (UH) band.
Thus, the effective two-band model of the superconducting phase transition
in cuprates retains from the complete set of $p$-$d$ bands inside the
CuO$\sb{2}$ plane precisely the ZR and UH bands and incorporates the overall
small effects coming from the other bands into the parameters of the model.
\par
Second, since $\Delta\sim 2W$, the resulting model Hamiltonian corresponds
to the \emph{strong correlation limit}.
This leads to considerable complications in the mathematical characterization
of the system excitations originating in inelastic interactions since the
kernel of the resulting integral representations of the corresponding Green
functions is not separable.
\medskip

{\centering\subsection{Role of the doping: cuprate families\label{sec:doping}}}

A high critical temperature superconductor does not simply exist as a well
defined stoichiometric structure.
It belongs to a \emph{family of cuprates\/} (e.g., LSCO, YBCO, etc.),
characterized by the occurrence of a characteristic stoichiometric
\emph{reference structure\/} and its specific \emph{kind of doping\/}
(either with \emph{holes}, giving rise to a \emph{hole doped cuprate family},
or with \emph{electrons}, giving rise to an \emph{electron doped cuprate
family}).
Within each family, the modification of the \emph{doping rate $\delta$\/}
results in drastic modifications of the physical properties.
\par
The reference structure is an \emph{insulator\/} characterized by a
\emph{strong antiferromagnetic exchange interaction\/} (the highest known
values of the antiferromagnetic exchange parameters occur in cuprates
\cite{Pl01}).
While the antiferromagnetic ordering is preserved at doping rates
$\delta < \delta\sp{1}\sb{cr}$, a doping range exists,
$\delta\sp{1}\sb{cr} < \delta < \delta\sp{2}\sb{cr}$,
at which a superconducting phase transition occurs with a doping dependent
measured onset critical temperatures,
$T\sb{c} = T\sb{c}(\delta)$.
At some \emph{optimum\/} doping rate,
$\delta = \delta\sb{opt}\in (\delta\sp{1}\sb{cr}, \delta\sp{2}\sb{cr})$,
the critical temperature $T\sb{c}$ reaches a maximum which allows the
characterization of the cuprate as a high critical temperature superconductor.
As a consequence, a given cuprate family exhibits a \emph{characteristic\/}
$(\delta ,T)$ \emph{phase diagram} (see, e.g.,
\cite{D03,Ka98,PWA00}).
\par
A consistent theoretical model has to give full account of the
$(\delta ,T)$ phase diagram, with correct reproduction of the antiferromagnetic
ordering at zero and low doping rates, of the characteristic properties
measured in the \emph{underdoped\/} ($\delta < \delta\sb{opt}$) and the
\emph{overdoped\/} ($\delta > \delta\sb{opt}$) regimes respectively and, of
course, to predict correctly the optimum doping rate
$\delta = \delta\sb{opt}$ at which the critical superconducting temperature
reaches its maximum across the family.
\medskip

{\centering\subsection{The effective spin lattice\label{sec:spinl}}}

The stoichiometric reference structure ($\delta=0$) of a cuprate family
is an insulator characterized by a very large gap ($\Delta\sim 2$ eV
\cite{PWA00,14Hb07})
between the two subbands lying nearest to the Fermi level.
As a consequence, its electron (hole) states are
\emph{frozen\/} at the nodes~$i$ of the two-dimensional regular lattice
defined by the positions of the copper ions inside the CuO$\sb{2}$ plane.
\par
The essentials of the behaviour of the system are preserved if the actual
CuO$\sb{2}$ lattice is replaced by an \emph{effective two-dimensional spin
lattice\/} having the spins placed at the copper sites inside the
CuO$\sb{2}$ plane.
Three consequences are immediate for the characterization of the effective
spin lattice. First, the \emph{spin lattice constants}, $a\sb{x}$ and
$a\sb{y}$, are given by the lattice constants of the physical
CuO$\sb{2}$ plane.
Second, since the reference physical structure shows antiferromagnetic
ordering, any pair of two first order neighbouring spins within the reference
effective spin lattice have the spins ordered in \emph{opposite directions}.
Third, there are four such possible states at each lattice site $i$
in the effective spin lattice:
$|0\rangle$ (vacuum), $|\sigma\rangle = \vert \! \! \uparrow \rangle$ and
$|\bar\sigma\rangle = \vert \! \! \downarrow \rangle$ (single particle spin
states inside the hole subband), and
$|2\rangle = \vert \! \! \uparrow \downarrow \rangle$ (singlet state in the
singlet subband).
\par
The \emph{doping\/} of the electron states within the physical
CuO$\sb{2}$ plane is equivalent to the creation of \emph{defects\/} inside
the spin lattice by means of either \emph{spin vacancies\/} and/or
\emph{singlet states}.
As a consequence of the occurrence of spin defects, the spin lattice
ceases to be frozen: \emph{hopping conduction\/} arises.
\medskip

{\centering\subsection{Hubbard operator description of hopping
\label{sec:HO}}}

The fact that the single $d$ electron states are tightly-bound at the sites
of the copper ions within the CuO$\sb{2}$ lattice accounts for the unusually
low conduction rate in the doped cuprates.
As a consequence, adequate description of the hopping processes between the
effective lattice spin sites is got
\cite{Pl95}
in terms of the \emph{Hubbard operators\/} (HOs)
\cite{Hub64},
  $X\sb{i}\sp{\alpha \beta} = |i\alpha\rangle \langle i\beta|$,
where $|\alpha\rangle$ and $|\beta\rangle$ denote the initial, respectively
final spin states at the spin lattice site $i$.
\par
At every spin lattice site $i$ the Hubbard operator multiplication rule holds
\begin{equation}
  X\sb{i}\sp{\alpha \beta}X\sb{i}\sp{\gamma\eta} = \delta \sb{\beta\gamma}
   X\sb{i}\sp{\alpha\eta},
\label{eq:multH}
\end{equation}
together with the completeness relation which secures the rigorous fulfillment
of the Pauli exclusion principle,
\begin{equation}
X\sb{i}\sp{00} + X\sb{i}\sp{\sigma\sigma} +
   X\sb{i}\sp{\bar{\sigma}\bar{\sigma}} + X\sb{i}\sp{22} = 1.
\label{eq:complH}
\end{equation}
\par
The single spin state creation/annihilation in a subband are described by
\emph{fermionic\/} HOs, while the singlet creation/annihilation, spin or
charge densities, particle numbers, are described by \emph{bosonic\/} HOs.
Therefore, the Hubbard operator algebra is defined both in terms of
anticommutation and commutation relations.
For a pair of fermionic HOs, the anticommutator rule holds
\begin{equation}
\{X\sb{i}\sp{\alpha \beta}, X\sb{j}\sp{\gamma\eta}\} = \delta\sb{ij}
           (\delta \sb{\beta\gamma}X\sb{i}\sp{\alpha\eta} +
            \delta \sb{\eta\alpha}X\sb{i}\sp{\gamma\beta}),
\label{eq:anticH}
\end{equation}
whereas, if one or both HOs are bosonic, the commutation rule holds
\begin{equation}
[X\sb{i}\sp{\alpha \beta}, X\sb{j}\sp{\gamma\eta}] =
   \delta\sb{ij}(\delta \sb{\beta\gamma}X\sb{i}\sp{\alpha\eta} -
   \delta \sb{\eta\alpha}X\sb{i}\sp{\gamma\beta}).
\label{eq:comutH}
\end{equation}
Since there are two kinds of inhomogeneities (vacancies and singlets)
introduced by the doping inside the spin lattice, two fundamentally different
hopping conduction processes will happen: \emph{fermion hopping\/} (of
single spins to vacancies inside the spin lattice or single spin interband
transitions) and \emph{boson hopping\/} (of singlet spin pairs to
vacancies inside the spin lattice).
\medskip

{\centering\subsection{Hopping conduction neighbourhood of a spin lattice site
\label{sec:hopn}}}

Projection techniques based on cell-cluster perturbation theory
\cite{Feiner96a}-\cite{Yushankhai97}
showed that the relative intensity of the hopping process relating the sites
$i$ and $m$ of the spin lattice is determined by the non-vanishing Wannier
coefficient $\nu\sb{im}$ following from the overlap of the wave functions
of the $d$-copper and $p$-oxygen states.
The coefficients $\nu\sb{im}$ show (non-exponential) decrease with the distance
$r\sb{im} = \vert{\bf r}\sb{m} - {\bf r}\sb{i}\vert$ inbetween the sites
$i$ and $m$.
Significantly different from zero are the Wannier coefficients within the
first three coordination spheres around a given reference site $i$.
An instance of typical values is
\cite{Pl95,PO07}:
for the nearest neighbouring (nn) $m$-sites (the first coordination sphere),
$\nu\sb{im}\sim\nu\sb{1} = 0.14$; for the next nearest neighbouring (nnn)
$m$-sites (the second coordination sphere),
$\nu\sb{im}\sim\nu\sb{2} = - 0.13\, \nu\sb{1}$, while for the $m$-sites located
at the third coordination sphere, $\nu\sb{im}\sim\nu\sb{3} = 0.16\, \nu\sb{1}$.
\par
The scrutiny of the hopping part of the Hamiltonian of the effective two-band
Hubbard model
\cite{Pl95}
showed
\cite{AA07}
(see also
\cite{RJP08})
that a hopping conduction neighbourhood of a given spin lattice site $i$ can
be defined in terms of the Hubbard 1-form of labels
$(\alpha\beta , \gamma\eta)$,
\begin{equation}
\tau\sb{1,i}\sp{\alpha \beta , \gamma \eta} = \sum\sb{m\neq i}
    \nu\sb{im}X\sb{i}\sp{\alpha \beta} X\sb{m}\sp{\gamma\eta}\ , \quad\quad
    \Big( \tau\sb{1,i}\sp{\alpha \beta , \gamma \eta}\Big)\sp{\dagger} =
    - \tau\sb{1,i}\sp{\beta\alpha , \eta\gamma}\ .
\label{eq:taup}
\end{equation}
The actual labels $(\alpha\beta , \gamma\eta)$ are defined by the available
in-band or inter-band transitions within the effective spin lattice.
\medskip

{\centering\subsection{Finite hopping induced energy correction effects
\label{sec:hopd}}}
The creation of defects inside the spin lattice by doping has two kinds of
consequences on the normal state of a cuprate. First, the emergence of
hopping processes result in the occurrence of hopping conductivity.
Second, since the hopping causes finite modifications of a small fraction
of the electron states in the neighboorhoud of the Fermi surface, the
correlations stemming from hopping induce \emph{finite\/} corrections
to the energy spectrum of the compound.
The fulfillment of this condition by the mean field energy spectrum which
follows from the effective two-band Hubbard model will be shown in
section~\ref{sec:mhH} to ask for a non-trivial modification of the hopping
part of its Hamiltonian.
\medskip

{\centering\subsection{Unconventional anomalous pairing in cuprates
\label{sec:upair}}}
Experimental measurements resulting in inferences on the pairing mechanism
in cuprates concern the charge and the spin of the superconducting current
carriers, as well as the phase of the energy gap $\Delta({\bf k})$.
Until the discovery of the cuprates, it was known that the anomalous pairing
yielding zero spin and charge $2e$ Cooper pairs in the superconducting
phase stems from the lattice phonon mediated interaction between conduction
electron pairs in metallic samples.
This $s$-wave pairing mechanism, which results in a gap function
$\Delta({\bf k})$ that preserves the symmetry of the Fermi surface of the
compound, was initially assumed in many papers to work in the cuprates as well.
\par
Flux quanta measurements in cuprates at temperatures lower than $T\sb{c}$
evidenced that the superconducting current is carried by electron pairs
having the charge $2e$
\cite{Gough87},
while the small but significant drop off of the Cu Knight shift below $T\sb{c}$
\cite{05Hb07}
pointed to singlet pairing in cuprates.
\par
Soon after the discovery of the cuprates, P.W.~Anderson
\cite{PWA87}
assumed that the pairing mechanism should be unconventional.
For the time being, there are several classes of models the starting
hypotheses of which predict the occurrence of a $d\sb{x\sp{2}-y\sp{2}}$-pairing
mechanism.
The two-dimensional two-band Hubbard model results in a static
$d\sb{x\sp{2}-y\sp{2}}$-pairing exchange mechanism (see
\cite{Pl03}
and
\cite{AA07}).
The inelastic correlation effects beyond the GMFA solution add a
$d\sb{x\sp{2}-y\sp{2}}$-pairing spin fluctuation mechanism as well
\cite{Pl03}.
\par
This $d$-pairing mechanism results in a gap function $\Delta({\bf k})$ the
symmetry of which is lower than that of the Fermi surface.
Phase-sensitive experiments, measuring the phase shifts in a dc SQUID
involving a corner Josephson junction
\cite{vH95},
or detecting half-flux quanta in a frustrated geometry
\cite{TsK00},
revealed that the occurrence of a robust $d\sb{x\sp{2}-y\sp{2}}$-pairing
is a common feature of both the hole-doped and electron-doped cuprates.
In hole-doped cuprates like YBCO, showing a small orthorhombic distortion
from the tetragonal reference lattice in the CuO$\sb{2}$ plane, a small
additional $s$-wave component was also evidenced.
\par
The phase-sensitive Andreev--Saint-James spectroscopy data
\cite{Deutscher05}
confirmed these findings.
\medskip

{\centering\subsection{Superconducting phase kinetic energy minimization
\label{sec:kin}}}
The occurrence of the superconducting phase below some critical temperature
$T\sb{c}$ happens as a result of the minimization of the total energy of the
system by the correlated spin configuration of Cooper pairs to a value which
is lower as compared to that of the normal Fermi-liquid state.
\par
In the conventional superconductors, the electron-phonon interaction mechanism,
which yields, below $T\sb{c}$, ordered Cooper pairs within a manifold of the
electron states lying near the Fermi level, results in significant minimization
of the potential energy of the system.
\par
Experimental data concerning the energy distribution of the superconducting
phase in cuprates
\cite{Science2002,Hirsch2002}
have shown that this associates the minimization of the kinetic energy of
tightly bound pairs of electron states.
Therefore, the anomalous pairing within a model consistent with these data
is to involve essentially the kinetic energy (i.e., hopping), while vanishing
or small correction contributions coming from the potential energy terms.
\par
The anomalous pairing within the two-dimensional two-band Hubbard model
stems mainly from the hopping part of the model Hamiltonian
\cite{Pl03,AA07},
hence it is of kinetic origin.
\medskip

{\centering\subsection{Spin-charge separation
\label{sec:s-c}}}
The spin-charge separation in cuprates, was shown by P.W.~Anderson
\cite{PWA00}
to provide natural understanding of the anomalous cuprate behaviour occurring
in all the four distinct phases of a $(\delta ,T)$ phase diagram: the
normal metallic phase, the pseudogap state separated from the previous one
by the temperature $T\sp{*} = T\sp{*}(\delta)$, the $d$-wave superconducting
phase, and the antiferromagnetically ordered phase which occurs at vanishing
doping or in the underdoped regime.
\par
As recently discussed in
\cite{RJP08,JOAM08},
the spin-charge separation is recovered in the GMFA rigorous solution of the
two-band Hubbard model
\cite{AA07}
as a consequence of the fact that the spin-charge correlation matrix elements
vanish exactly.
\bigskip
\par
{\centering\section{STANDARD MODEL HAMILTONIAN\label{sec:H}}}
Using~(\ref{eq:taup}), the Hamiltonian of the effective two-band Hubbard model
\cite{Pl95}
can be rewritten in the locally manifest Hermitian form
\begin{equation}
      H = H\sb{0} + H\sb{h}\ ,
\label{eq:H}
\end{equation}
with the single particle contribution
\begin{eqnarray}
      H\sb{0}&\! \! \! \! =\! \! \! \! &\sum\sb{i} h\sb{0,i}\ ,
            \quad\quad h\sb{0,i}\sp{\dagger} = h\sb{0,i}\ ,
\nonumber\\
      h\sb{0,i}&\! \! \! \! =\! \! \! \! &
      E\sb{1} \sum\sb{\sigma} X\sb{i}\sp{\sigma \sigma} \! +\!
      E\sb{2} X\sb{i}\sp{22} \ ,
\label{eq:H0}
\end{eqnarray}
and the hopping contribution
\begin{eqnarray}
      H\sb{h}&\! \! \! \! =\! \! \! \! &\sum\sb{i} h\sb{h,i}\ ,
            \quad\quad h\sb{h,i}\sp{\dagger} = h\sb{h,i}\ ,
\nonumber\\
      h\sb{h,i}&\! \! \! \! =\! \! \! \! &
      \frac{\mathcal{K}\sb{11}}{2}\sum\sb{\sigma}
      \Big(\tau\sb{1,i}\sp{\sigma 0,0\sigma}\! \! -\!
           \tau\sb{1,i}\sp{0\sigma ,\sigma 0}\Big)\! +\!
      \frac{\mathcal{K}\sb{22}}{2}\sum\sb{\sigma}
      \Big(\tau\sb{1,i}\sp{2\sigma ,\sigma 2}\! \! -\!
           \tau\sb{1,i}\sp{\sigma 2,2\sigma}\Big)\! +\!
\nonumber\\
           &\! \! \! \! +\! \! \! \! &
   \frac{\mathcal{K}\sb{21}}{2}\sum\sb{\sigma}2\sigma
       \Big[ \Big(\tau\sb{1,i}\sp{2\bar{\sigma},0\sigma}\! \! -\!
                  \tau\sb{1,i}\sp{0\sigma ,2\bar{\sigma}}\Big) \! +\!
             \Big(\tau\sb{1,i}\sp{\sigma 0, \bar{\sigma}2}\! \! -\!
                  \tau\sb{1,i}\sp{\bar{\sigma}2,\sigma 0}\Big)
       \Big] \ .
\label{eq:Hh}
\end{eqnarray}
In these equations the summation label $i$ runs over the sites of an
infinite two-dimensional lattice with the lattice constants $a\sb{x}$
and $a\sb{y}$ respectively defined by the crystal structure of the cuprate.
The spin projection values in the sums over $\sigma$ are
$\sigma = \pm 1/2$, $\bar\sigma = -\sigma$.
\par
In~(\ref{eq:H}), $E\sb{1} = \tilde{\varepsilon \sb{d}} - \mu$ denotes the
hole subband energy for the renormalized energy
$\tilde{\varepsilon \sb{d}}$ of a $d$-hole and the chemical potential $\mu$.
The energy parameter of the singlet subband is
$E\sb{2} = 2E\sb{1} + \Delta$, where
$\Delta\approx\Delta\sb{pd} = \varepsilon\sb{p}-\varepsilon\sb{d}\simeq 2 eV$
is an effective Coulomb energy $U\sb{\rm eff}$ corresponding to the difference
between the two energy levels of the model.
The hopping energy parameters $\mathcal{K}\sb{ab} = 2 t\sb{pd}K\sb{ab}$
($a, b = 1, 2$)
depend on $t\sb{pd}$, the hopping $p$-$d$ integral, and on energy band
dependent form factors $K\sb{ab}$. The label $1$ points to the
hole subband, while $2$ to the singlet subband.
Inband ($\mathcal{K}\sb{11}, \mathcal{K}\sb{22}$) and interband
($\mathcal{K}\sb{21} = \mathcal{K}\sb{12}$) processes are present.
\par
The quasi-particle spectrum and the superconducting pairing within the
Hamiltonian~(\ref{eq:H}) are obtained
\cite{Pl97,Pl03}
by the equation of motion technique for the retarded and advanced two-time
$4\times 4$ GF matrices in the (${\bf r}, t$)-representation, which represent
a single matrix in the (${\bf q}, \omega$)-representation.
The GMFA solution of this matrix is summarized in the next section.
\bigskip
\par
{\centering\section{MEAN-FIELD APPROXIMATION\label{sec:gmfa}}}
The Green function matrices of the model define space-time correlations for
the four-component Nambu column operator
\cite{Pl97,Pl03}
\begin{equation}
  \hat X\sb{i\sigma}=(X\sb{i}\sp{\sigma 2}\,\,
  X\sb{i}\sp{0\bar\sigma}\,\, X\sb{i}\sp{2\bar\sigma}\,\,
  X\sb{i}\sp{\sigma 0})\sp{\top}
\label{eq:nambu}
\end{equation}
and its adjoint operator
  $\hat X\sb{j\sigma}\sp{\dagger} = (X\sb{j}\sp{2\sigma}\,\,
  X\sb{j}\sp{\bar\sigma 0}\,\, X\sb{j}\sp{\bar\sigma 2}\,\,
  X\sb{j}\sp{0\sigma})$.
In~(\ref{eq:nambu}), the superscript $\top$ denotes the transposition.
\par
The retarded GF matrix is written, in Zubarev notation
\cite{Zub60},
as follows
\begin{equation}
         \tilde G\sp{(r)}\sb{ij\sigma}(t-t')  =
    \langle\langle \hat X\sb{i\sigma}(t)\! \mid \!
    \hat X\sb{j\sigma}\sp{\dagger}(t')\rangle\rangle =
    -{\rm i}\theta (t-t')\langle \{\hat X\sb{i\sigma}(t),
    \hat X\sb{j\sigma}\sp{\dagger}(t')\}\rangle ,
\label{eq:GF}
\end{equation}
where $\langle \cdots \rangle$ denotes statistical average over
Gibbs grand canonical ensemble.
\par
The advanced GF matrix replaces in~(\ref{eq:GF}) the temporal factor by
${\rm i}\theta (t'-t)$.
\par
The GF matrix in the (${\bf r}, \omega$)-representation is related to
the GF matrix in the (${\bf r}, t$)-representation by the non-unitary
Fourier transform,
\begin{equation}
    \tilde G\sb{ij\sigma}(t-t')  = \frac{1}{2\pi}
    \int\limits_{-\infty}^{+\infty}\tilde G\sb{ij\sigma}(\omega)
    \; {\rm e}\sp{-{\rm i}\omega (t-t\sp{\prime})}\, {\rm d}\omega \; ,
\label{eq:t2om}
\end{equation}
where the superscripts $(r)$ and $(a)$ have been omitted.
\par
The analytic continuations of the retarded and advanced Green functions
in the complex energy $\omega$-plane define a single complex function, denoted
$\tilde G\sb{ij\sigma}(\omega)$, with cuts (jumps) along the real energy axis.
\par
The energy spectrum of the Hamiltonian~(\ref{eq:H})
is solved in the reciprocal space. The GF matrix in this
(${\bf q}, \omega$)-representation is related to the GF matrix in
(${\bf r}, \omega$)-representation by the non-unitary discrete Fourier
transform
\begin{equation}
    \tilde G\sb{ij\sigma}(\omega) =
    \frac{1}{N}\sum\sb{\bf q} {\rm e}\sp{-{\rm i}{\bf q}\;
        ({\bf r}\sb{j} - {\bf r}\sb{i})} \;
    \tilde G\sb{\sigma}({\bf q}, \omega).
\label{eq:r2q}
\end{equation}
\par
For an elemental Green function of labels $(\alpha\beta,\gamma\eta)$,
we use the notation
$\langle\langle X\sb{i}\sp{\alpha\beta}(t)|
                X\sb{j}\sp{\gamma\eta}(t')\rangle\rangle$
in the (${\bf r}, t$)-representation
and, similarly,
$\langle\langle X\sb{i}\sp{\alpha\beta}|
                X\sb{j}\sp{\gamma\eta}\rangle\rangle\sb{\omega}$
(assuming Hubbard operators at $t=0$),
in the (${\bf r}, \omega$)-representation.
In the (${\bf q}, \omega$)-representation, it is convenient to use the
notation $G\sp{\alpha\beta,\gamma\eta}({\bf q}, \omega)$.
\par
We shall consider henceforth the GMFA-GF,
$\tilde G\sb{\sigma}\sp{0}({\bf q}, \omega)$ in the form
\cite{AA07},
\begin{eqnarray}
  \tilde G\sp{0}\sb{\sigma }({\bf q},\omega) = \tilde \chi\;
    \Bigl[ \tilde \chi \omega - \tilde \mathcal{A}\sb{\sigma}({\bf q})
    \Bigr] \sp{-1} \tilde \chi\; ; \! \! \! &&\! \! \!
   \tilde\chi =
     \langle \{\hat X\sb{i\sigma},\hat X\sb{i\sigma}\sp{\dagger}\}\rangle ;
\label{eq:G0A}\\
    \tilde \mathcal{A}\sb{\sigma}({\bf q}) =
    \sum\sb{{\bf r}\sb{ij}} {\rm e}\sp{i{\bf q}
        \cdot{\bf r}\sb{ij}}
    \tilde \mathcal{A}\sb{ij\sigma}\; ; \ \
      {\bf r}\sb{ij} = {\bf r}\sb{j} - {\bf r}\sb{i}\; ; \! \! \! &&\! \! \!
  \tilde \mathcal{A}\sb{ij\sigma} = \langle \{ [\hat X\sb{i\sigma}, H],
    \hat X\sb{j\sigma}\sp{\dagger} \} \rangle \, .
\label{eq:Aij}
\end{eqnarray}
\par
Two kinds of particle number operators, related to the singlet subband,
\begin{equation}
   n\sb{i\sigma} = X\sb{i}\sp{\bar\sigma\bar\sigma}+X\sb{i}\sp{22}\, , \quad
   n\sb{i\bar\sigma} = X\sb{i}\sp{\sigma\sigma} + X\sb{i}\sp{22}\, , \quad
    N\sb{i} = n\sb{i\sigma} + n\sb{i\bar\sigma}\, ,
\label{eq:opni}
\end{equation}
and to the hole subband respectively,
\begin{equation}
   n\sb{i\sigma}\sp{h} = X\sb{i}\sp{\sigma\sigma} + X\sb{i}\sp{00}\, , \quad
   n\sb{i\bar\sigma}\sp{h} = X\sb{i}\sp{\bar\sigma\bar\sigma} +
                             X\sb{i}\sp{00}\, , \quad
   N\sb{i}\sp{h} = n\sb{i\sigma}\sp{h} + n\sb{i\bar\sigma}\sp{h}\, .
\label{eq:opnih}
\end{equation}
can be defined. The completeness relation implies
   $n\sb{i\sigma} + n\sb{i\sigma}\sp{h} =
   n\sb{i\bar\sigma} + n\sb{i\bar\sigma}\sp{h} = 1\, .$
\par
The $\tilde\chi$ matrix in~(\ref{eq:G0A}) is diagonal,
\begin{eqnarray}
  \tilde\chi &=& \left(
    \begin{array}{cc}
      \hat{\chi} & \hat{0}\\
         \hat{0} & \hat{\chi}
    \end{array}
              \right), \quad
  \hat{\chi} = \left(
    \begin{array}{cc}
      \chi\sb{2} & 0\\
               0 & \chi\sb{1}
    \end{array}
              \right), \quad
  \hat{0} = \left(
    \begin{array}{cc}
      0 & 0\\
      0 & 0
    \end{array}
              \right) ,
\label{eq:chicalc}
\end{eqnarray}
where $\chi\sb{2}$ and $\chi\sb{1}$ denote spin and site independent averages,
\begin{equation}
   \chi\sb{2} = \langle n\sb{i\sigma}\rangle =
   \langle n\sb{i\bar\sigma}\rangle ;  \quad 
   \chi\sb{1} = \langle n\sb{i\sigma}\sp{h}\rangle =
   \langle n\sb{i\bar\sigma}\sp{h}\rangle = 1 - \chi\sb{2}\; .
\label{eq:chi1}
\end{equation}
In terms of the doping rate $\delta$, it results that in the
hole-doped cuprates,
\begin{equation}
   \chi\sb{2} = \delta , \quad
   \chi\sb{1} = 1 - \delta ,
\label{eq:chihd}
\end{equation}
while in the electron-doped cuprates,
\begin{equation}
   \chi\sb{1} = \delta , \quad
   \chi\sb{2} = 1 - \delta .
\label{eq:chied}
\end{equation}
\par
To understand the consequences following for the energy spectrum, we rewrite
the GMFA-GF~(\ref{eq:G0A}) in the algebraically equivalent form
\begin{equation}
  \tilde G\sp{0}\sb{\sigma }({\bf q},\omega) = \tilde \chi\sp{1/2}
  \Bigl[ I\omega - \tilde\mathcal{E}\sb{\sigma}({\bf q})
  \Bigr] \sp{-1} \tilde \chi\sp{1/2}\;,
\label{eq:G0qonorm}
\end{equation}
where $I$ denotes the $4\times 4$ unit matrix, while
$\tilde\mathcal{E}\sb{\sigma}({\bf q})$ is the Hermitian matrix
\begin{equation}
  \tilde\mathcal{E}\sb{\sigma}({\bf q}) =
  \tilde \chi\sp{-1/2}\tilde\mathcal{A}\sb{\sigma}({\bf q})
  \tilde \chi\sp{-1/2}\;.
\label{eq:Esqdef}
\end{equation}
The GMFA spectrum of the model Hamiltonian~(\ref{eq:H}) is therefore given by
the eigenvalues of $\tilde \mathcal{E}\sb{\sigma}({\bf q})$.
Using the results reported in
\cite{AA07},
we get 
\begin{equation}
  \tilde \mathcal{E}\sb{\sigma}({\bf q}) =
  \left( \begin{array}{cc}
    \hat E\sb{\sigma}({\bf q})& \hat \Phi\sb{\sigma}({\bf q})\\
    (\hat \Phi\sb{\sigma}({\bf q}))\sp{\dagger} &
      - (\hat {E}\sb{\bar\sigma}({\bf q}))\sp{\top}
  \end{array} \right) .
\label{eq:Eqs}
\end{equation}
\par
The normal correlations contribute the $2\times 2$ matrices,
\begin{equation}
  \hat E\sb{\sigma}({\bf q}) =
  \left( \begin{array}{cc}
    C\sb{22} & 2\sigma C\sb{21}\\
    2\sigma C\sb{21}\sp{*}&C\sb{11}
  \end{array} \right) , \quad
  - (\hat {E}\sb{\bar\sigma}({\bf q}))\sp{\top} =
  \left( \begin{array}{cc}
    - C\sb{22} & 2\sigma C\sb{21}\sp{*}\\
    2\sigma C\sb{21}& -C\sb{11}
  \end{array} \right) ,
\label{eq:Eqscn}
\end{equation}
with the distinct matrix elements
\begin{eqnarray}
   C\sb{22}\! \! &\equiv&\! \! C\sb{22}({\bf q}) = (E\sb{1}+\Delta) +
             [a\sb{22} + d\sb{22}({\bf q})]/\chi\sb{2}
\label{eq:C22}\\
   C\sb{11}\! \! &\equiv&\! \! C\sb{11}({\bf q}) = E\sb{1} +
             [a\sb{22} + d\sb{11}({\bf q})]/\chi\sb{1}
\label{eq:C11}\\
   C\sb{21}\! \! &\equiv&\! \! C\sb{21}({\bf q}) =
             [a\sb{21} + d\sb{21}({\bf q})]/\sqrt{\chi\sb{1}\chi\sb{2}}
\label{eq:C21}
\end{eqnarray}
Here $a\sb{mn}$ and $d\sb{mn}({\bf q})$ denote respectively
the one-site and two-site contributions to the matrix
$\tilde \mathcal{A}\sb{\sigma}({\bf q})$ coming from the \emph{hopping
Hamiltonian}~(\ref{eq:Hh}):
\begin{eqnarray}
    a\sb{22}\! \! &=&\! \! \mathcal{K}\sb{11}
             \langle\tau\sb{1}\sp{0\bar\sigma, \bar\sigma 0}\rangle -
            \mathcal{K}\sb{22}\langle\tau\sb{1}\sp{\sigma 2, 2\sigma}\rangle ,
\label{eq:a22}\\
  a\sb{21}\! \! &=&\! \! (\mathcal{K}\sb{11}-\mathcal{K}\sb{22})\cdot 2\sigma
                \langle\tau\sb{1}\sp{\sigma 2, \bar\sigma 0}\rangle +
  \mathcal{K}\sb{21}(\langle\tau\sb{1}\sp{0\bar\sigma, \bar\sigma 0}\rangle-
                \langle\tau\sb{1}\sp{\sigma 2, 2\sigma}\rangle),
\label{eq:a21}\\
  d\sb{mn}({\bf q})\! \! &=&\! \! \mathcal{K}\sb{mn}\sum\sb{\alpha=1}\sp{3}
      \nu\sb{\alpha}\gamma\sb{\alpha}({\bf q})[\chi\sb{\alpha}\sp{S} +
      (-1)\sp{m+n}\chi\sb{m}\chi\sb{n}] +
      \frac{1}{2} J\sb{mn} \chi\sp{\rm s-h}({\bf q}).
\label{eq:dabq}
\end{eqnarray}
Here and in what follows, $\langle\tau\sb{1}\sp{\lambda\mu, \nu\varphi}\rangle$
denotes the site-independent average of the Hubbard 1-form~(\ref{eq:taup}),
\begin{equation}
   \langle\tau\sb{1}\sp{\lambda\mu, \nu\varphi}\rangle =
   \sum\sb{\alpha=1}\sp{3}\nu\sb{\alpha}\cdot
   \frac{1}{N}\sum\sb{\bf q}\langle X\sp{\lambda\mu}
     X\sp{\nu\varphi}\rangle\sb{\bf q}\gamma\sb{\alpha}({\bf q})
\label{eq:tau1o}
\end{equation}
for all the label sets $(\lambda\mu, \nu\varphi)$ of interest.
Further,
\begin{equation}
        \langle X\sp{\lambda\mu}X\sp{\nu\varphi}\rangle\sb{\bf q} =
        \frac{i}{2\pi}\int\limits_{-\infty}^{+\infty}
        \frac{{\rm d}\omega}{1+{\rm e}\sp{-\beta\omega}}
        \Big[G\sp{\lambda\mu, \nu\varphi}({\bf q}, \omega+i\varepsilon) -
        G\sp{\lambda\mu, \nu\varphi}({\bf q}, \omega-i\varepsilon)\Big] ,
\label{eq:SpTq}
\end{equation}
while the quantities $\gamma\sb{\alpha}({\bf q})$ denote the nn
     $(\alpha=1)$, nnn $(\alpha=2)$, and third neighbour $(\alpha=3)$
     geometrical form factors,
   $\gamma\sb{1}({\bf q}) = 2 [\cos(q\sb{x}a\sb{x}) + \cos(q\sb{y}a\sb{y})]$,
   $\gamma\sb{2}({\bf q}) = 4 \cos(q\sb{x}a\sb{x}) \cos(q\sb{y}a\sb{y})$,
   $\gamma\sb{3}({\bf q}) = 2 [\cos(2q\sb{x}a\sb{x}) + \cos(2q\sb{y}a\sb{y})]$.
\par
In equation~(\ref{eq:dabq}),
$\{ \chi\sb{\alpha}\sp{S}\, \vert\, \alpha = 1, 2, 3 \}$
denote the nn, nnn, and third coordination sphere parameters respectively
coming from the phenomenological representation of the spin-spin correlation
function $\langle {\bf S}\sb{i}{\bf S}\sb{j} \rangle$.
\par
The exchange energy parameters are given by
\begin{equation}
     J\sb{mn} = 4\mathcal{K}\sb{mn} \mathcal{K}\sb{21}/\Delta , \quad
          \{mn\} \in \{22, 11, 21\} .
\label{eq:Jab}
\end{equation}
\par
Finally, the singlet hopping contribution $\chi\sp{\rm s-h}({\bf q})$
is given by
\begin{equation}
       \chi\sp{\rm s-h} ({\bf q}) =
       \sum\sb{\alpha=1}\sp{3}\nu\sb{\alpha}\sp{2}\cdot
       \frac{1}{N}\sum\sb{\bf k}\Xi\sb{\bf k}
       \gamma\sb{\alpha}({\bf q}-{\bf k})
\label{eq:shhd1}
\end{equation}
where $\Xi\sb{\bf k} = 2\sigma\langle X\sp{\sigma 2}
         X\sp{\bar\sigma 0}\rangle\sb{\bf k}$,
while $\Xi\sb{\bf k} = 2\sigma\langle X\sp{0\bar\sigma}
         X\sp{2\sigma}\rangle\sb{\bf k}$ for hole-doped and
electron-doped cuprates respectively, with averages defined by~(\ref{eq:SpTq}).
\par
The anomalous correlations contribute to~(\ref{eq:Eqs}) the $2\times 2$
matrices,
\begin{equation}
  \hat \Phi\sb{\sigma}({\bf q}) =
  \left(
   \begin{array}{cc}
    - 2\sigma T\sb{2} & T\sb{21}\\
    - T\sb{21}        & 2\sigma T\sb{1}
   \end{array}
  \right) ; \quad
  (\hat \Phi\sb{\sigma}({\bf q}))\sp{\dagger} =
  \left(
   \begin{array}{cc}
    - 2\sigma T\sb{2}\sp{*} & - T\sb{21}\sp{*}\\
    T\sb{21}\sp{*}          & 2\sigma T\sb{1}\sp{*}
   \end{array}
  \right)
\label{eq:Fqs}
\end{equation}
where
\begin{eqnarray}
   T\sb{2}\! \! &\equiv&\! \! T\sb{2}({\bf q}) =
      [\mathcal{K}\sb{22}b\sb{1} + (1-\delta)\xi\sb{1}b\sb{2}({\bf q}) +
      \delta\xi\sb{1}b\sb{3}({\bf q})]/\chi\sb{2}
\label{eq:T2}\\
   T\sb{1}\! \! &\equiv&\! \! T\sb{1}({\bf q}) =
      [\mathcal{K}\sb{11}b\sb{1} + (1-\delta)\xi\sb{1}b\sb{2}({\bf q}) +
      \delta\xi\sb{1}b\sb{3}({\bf q})]/\chi\sb{1}
\label{eq:T1}\\
   T\sb{21}\! \! &\equiv&\! \! T\sb{21}({\bf q}) =
      [\mathcal{K}\sb{21}b\sb{1} + (1-\delta)\xi\sb{2}b\sb{2}({\bf q}) +
      \delta\xi\sb{2}b\sb{3}({\bf q})]/\sqrt{\chi\sb{1}\chi\sb{2}}
\label{eq:T21}
\end{eqnarray}
with the two-site exchange energies
$\xi\sb{1} = J\sb{21}$ and $\xi\sb{2} = (J\sb{11} + J\sb{22})/2$.
\par
The anomalous one-site pairing matrix elements are given by Hubbard 1-form
averages~(\ref{eq:tau1o}),
\begin{equation}
   b\sb{1} =
      \sum\sb{\sigma}2\sigma
         \langle\tau\sb{1}\sp{\sigma 2,\bar\sigma 2}\rangle =
      \sum\sb{\sigma}2\sigma
         \langle\tau\sb{1}\sp{0\bar\sigma ,0\sigma}\rangle\, ,
\label{eq:tauan1}
\end{equation}
where the first expression is to be used for hole-doped cuprates, while
the second one for electron-doped cuprates.
\par
The anomalous two-site pairing matrix elements following from the reduction
to localized Cooper pairs
\cite{AA07}
are
\begin{equation}
   b\sb{2}({\bf q}) =
       \sum\sb{\alpha=1}\sp{3}\nu\sb{\alpha}\sp{2}\cdot
       \frac{1}{N}\sum\sb{\bf k}\Pi\sb{\bf k}
       \gamma\sb{\alpha}({\bf q}-{\bf k})
\label{eq:phd1}
\end{equation}
where $\Pi\sb{\bf k} = 2\bar\sigma\langle X\sp{\sigma 2}
         X\sp{\bar\sigma 2}\rangle\sb{\bf k}$,
while $\Pi\sb{\bf k} = 2\sigma\langle X\sp{0 \bar\sigma}
         X\sp{0\sigma}\rangle\sb{\bf k}$  for hole-doped and
electron-doped cuprates respectively, with averages defined
in~(\ref{eq:SpTq}).
\par
The anomalous three-site pairing matrix elements following from the reduction
to localized Cooper pairs
\cite{AA07}
and the splitting of the three-site terms as done in
\cite{JOAM08}
are
\begin{equation}
   b\sb{3}({\bf q}) = \Pi\sb{\bf k}\sp{(3)}
       \sum\sb{\alpha=1}\sp{3}\nu\sb{\alpha}\cdot
       \gamma\sb{\alpha}({\bf q})
\label{eq:phd3}
\end{equation}
where $\Pi\sb{\bf k}\sp{(3)} = 2\bar\sigma
  \langle \tau\sb{1}\sp{\sigma 2,\bar\sigma 2}\rangle$,
while $\Pi\sb{\bf k}\sp{(3)} = 2\sigma
  \langle\tau\sb{1}\sp{0\bar\sigma ,0\sigma}\rangle$
for hole-doped and electron-doped cuprates respectively, with averages
defined in~(\ref{eq:tau1o}).
\bigskip
\par
{\centering\section{MODIFIED HOPPING HAMILTONIAN\label{sec:mhH}}}
The results derived in the previous section provide the rigorous GMFA-GF
solution for the energy matrix $\tilde\mathcal{E}\sb{\sigma}({\bf q})$ of
the effective Hamiltonian~(\ref{eq:H}).
These results have been derived assuming as starting hypotheses of the model
six out of the eleven features discussed in section~\ref{sec:essf}, namely
those listed in the subsections~\ref{sec:lams}--\ref{sec:cti} and
\ref{sec:spinl}--\ref{sec:hopn}.
\par
The features discussed in subsections~\ref{sec:upair} and~\ref{sec:kin}
are immediate consequences of the results reported in
\cite{Pl03,AA07}
and in the previous section.
The spin-charge separation (subsection~\ref{sec:s-c}) was shown in
\cite{RJP08,JOAM08}
to be a straightforward consequence of the exact vanishing of the spin-charge
correlation functions within the model.
\par
The features mentioned in subsections~\ref{sec:doping} and~\ref{sec:hopd}
ask for the \emph{finiteness\/} of all the terms of the matrix
$\tilde\mathcal{E}\sb{\sigma}({\bf q})$ at any value of the doping $\delta$,
in particular at vanishing doping, $\delta = 0$.
\par
From the equations~(\ref{eq:a22})--(\ref{eq:shhd1})
and~(\ref{eq:tauan1})--(\ref{eq:phd3}), it results that both the normal and
anomalous matrix elements coming from the hopping Hamiltonian~(\ref{eq:Hh})
are \emph{finite\/} in the limit of vanishing doping $\delta\rightarrow 0$.
\par
Corroborating this result with the values~(\ref{eq:chihd})
and~(\ref{eq:chied}) of the $\chi\sb{1}$ and $\chi\sb{2}$ parameters, from
the equations~(\ref{eq:C22})--(\ref{eq:C21}) and~(\ref{eq:T2})--(\ref{eq:T21})
it results that, in the hole-doped cuprates, the normal terms
$C\sb{22}({\bf q})$ and $C\sb{21}({\bf q})$, as well as the anomalous terms
$T\sb{2}({\bf q})$ and $T\sb{21}({\bf q})$ become \emph{infinite\/} in the
limit $\delta\rightarrow 0$ due to the vanishing denominator $\chi\sb{2}$.
Similarly, in the electron-doped cuprates, the normal terms
$C\sb{11}({\bf q})$ and $C\sb{21}({\bf q})$, as well as the anomalous terms
$T\sb{1}({\bf q})$ and $T\sb{21}({\bf q})$ become \emph{infinite\/} in the
same limit due to the vanishing denominator $\chi\sb{1}$.
\par
A simple remedy to this inconsistency of the standard Hamiltonian of the
model can be proposed from the scrutiny of the reduction process
resulting in the effective Hamiltonian~(\ref{eq:H}).
The derivation of the hopping parameters was done under the hypothesis of
occupied $d$-copper and $p$-oxygen states of interest.
However, under doping, part of these orbitals is empty and this fact is to
be reflected in the occurrence of an \emph{explicit\/} doping rate dependence
of the effective hopping parameters.
\par
The simplest way is to assume the renormalization, with a convenient factor
$\rho$, of the hopping Hamiltonian~(\ref{eq:Hh}) as a whole.
There are three possibilities to implement such a renormalization.
\par
The first is to assume $\rho = \delta$. Such a hypothesis would induce,
however, unphysical infinities in the theoretical limit $\delta\rightarrow 1$.
\par
The second possibility is to assume
$\rho = \min\{ \chi\sb{1}, \chi\sb{2} \}$, which would cure both the limits
$\delta\rightarrow 0$ and $\delta\rightarrow 1$, while inducing, however,
some peculiarities (turning points at $\delta = 0.5$) in the doping dependence
of both the normal and anomalous matrix elements entering
$\tilde\mathcal{E}\sb{\sigma}({\bf q})$.
\par
The third possibility, which results in \emph{smooth\/} dependence of the
matrix elements of $\tilde\mathcal{E}\sb{\sigma}({\bf q})$ on the doping rate
$\delta\in [0, 1]$ is to define $\rho = \chi\sb{1}\chi\sb{2}$, which can be
understood as simply assuming the site-pairs availability for fermion
hopping processes.
\par
Under this hypothesis, the normal matrix
elements~(\ref{eq:C22})--(\ref{eq:C21}) change to
\begin{eqnarray}
   C\sb{22}\! \! &\equiv&\! \! C\sb{22}({\bf q}) = (E\sb{1}+\Delta) +
             [a\sb{22} + d\sb{22}({\bf q})]\cdot\chi\sb{1}
\label{eq:C22m}\\
   C\sb{11}\! \! &\equiv&\! \! C\sb{11}({\bf q}) = E\sb{1} +
             [a\sb{22} + d\sb{11}({\bf q})]\cdot\chi\sb{2}
\label{eq:C11m}\\
   C\sb{21}\! \! &\equiv&\! \! C\sb{21}({\bf q}) =
             [a\sb{21} + d\sb{21}({\bf q})]\cdot
             \sqrt{\chi\sb{1}\chi\sb{2}},
\label{eq:C21m}
\end{eqnarray}
with the singlet hopping matrix element~(\ref{eq:shhd1}) replaced by
\begin{equation}
       \chi\sp{\rm s-h} ({\bf q}) = \chi\sb{1}\chi\sb{2}
       \sum\sb{\alpha=1}\sp{3}\nu\sb{\alpha}\sp{2}\cdot
       \frac{1}{N}\sum\sb{\bf k}\Xi\sb{\bf k}
       \gamma\sb{\alpha}({\bf q}-{\bf k}) .
\label{eq:shhdm}
\end{equation}
\par
The anomalous matrix elements~(\ref{eq:T2})--(\ref{eq:T21}) change now to
\begin{eqnarray}
   T\sb{2}\! \! &\equiv&\! \! T\sb{2}({\bf q}) =
      [\mathcal{K}\sb{22}b\sb{1} + (1-\delta)\xi\sb{1}b\sb{2}({\bf q}) +
      \delta\xi\sb{1}b\sb{3}({\bf q})]\cdot\chi\sb{1}
\label{eq:T2m}\\
   T\sb{1}\! \! &\equiv&\! \! T\sb{1}({\bf q}) =
      [\mathcal{K}\sb{11}b\sb{1} + (1-\delta)\xi\sb{1}b\sb{2}({\bf q}) +
      \delta\xi\sb{1}b\sb{3}({\bf q})]\cdot\chi\sb{2}
\label{eq:T1m}\\
   T\sb{21}\! \! &\equiv&\! \! T\sb{21}({\bf q}) =
      [\mathcal{K}\sb{21}b\sb{1} + (1-\delta)\xi\sb{2}b\sb{2}({\bf q}) +
      \delta\xi\sb{2}b\sb{3}({\bf q})]\cdot\sqrt{\chi\sb{1}\chi\sb{2}} ,
\label{eq:T21m}
\end{eqnarray}
with the isotropic one-site pairing matrix element $b\sb{1}$,
Eq.~(\ref{eq:tauan1}), left unchanged, the two-site Cooper pair
term~(\ref{eq:phd1}) replaced by
\begin{equation}
   b\sb{2}({\bf q}) = \chi\sb{1}\chi\sb{2}
       \sum\sb{\alpha=1}\sp{3}\nu\sb{\alpha}\sp{2}\cdot
       \frac{1}{N}\sum\sb{\bf k}\Pi\sb{\bf k}
       \gamma\sb{\alpha}({\bf q}-{\bf k})
\label{eq:phdm}
\end{equation}
and the effective three-site term contribution~(\ref{eq:phd3}) replaced by
\begin{equation}
   b\sb{3}({\bf q}) = \chi\sb{1}\chi\sb{2}\cdot\Pi\sb{\bf k}\sp{(3)}
       \sum\sb{\alpha=1}\sp{3}\nu\sb{\alpha}\cdot
       \gamma\sb{\alpha}({\bf q}) .
\label{eq:phd3m}
\end{equation}
\par
It is worthwhile to note that the crystallographic symmetry of the
CuO$\sb{2}$ lattice results in sizeable consequences on the relationships
among the anomalous pairing matrix elements $b\sb{1}$, $b\sb{2}({\bf q})$,
and $b\sb{3}({\bf q})$.
\par
In square CuO$\sb{2}$ lattices, where $a\sb{x} = a\sb{y}$ (e.g.~for
Bi-22(n-1)n, Tl-22(n-1)n, Hg-12(n-1)n, and Tl-12(n-1)n cuprate families
with n = 1, 2, 3), due to the existence of the $C\sb{4}$ rotation axis
symmetry, the anomalous averages 
$\langle\tau\sb{1}\sp{\sigma 2,\bar\sigma 2}\rangle$ and
$\langle\tau\sb{1}\sp{0\bar\sigma ,0\sigma}\rangle$ vanish identically.
Therefore the isotropic anomalous one-site term $b\sb{1}$,
Eq.~(\ref{eq:tauan1}), and the three-site term $b\sb{3}({\bf q})$,
Eq.~(\ref{eq:phd3m}), obtained after the splitting
\cite{RJP08,JOAM08}
of the three-site higher order correlation terms
(\cite{AA07},
section~6), equate to zero, such that the anomalous contributions~%
(\ref{eq:T2m})--(\ref{eq:T21m}) to the energy matrix~(\ref{eq:Eqs})
essentially reduce to the two-site terms retained in
\cite{AA07}.
\par
The rectangular CuO$\sb{2}$ lattices (e.g., for the cuprate family YBCO),
are very slightly different from square ones
($|a\sb{x} - a\sb{y}| \ll \min \{ a\sb{x}, a\sb{y}\}$).
Thus, after use of the $C\sb{4}$ rotation, small non-vanishing
values of the anomalous averages
$\langle\tau\sb{1}\sp{\sigma 2,\bar\sigma 2}\rangle$ and
$\langle\tau\sb{1}\sp{0\bar\sigma ,0\sigma}\rangle$ arise.
Then the terms $b\sb{1}$ and $b\sb{3}({\bf q})$
in~(\ref{eq:T2m})--(\ref{eq:T21m}) are small almost everywhere inside the
first Brillouin zone as compared to $b\sb{2}({\bf q})$.
Since the isotropic $b\sb{1}$ term points to the existence of an $s$-type
contribution to the anomalous pairing, qualitative agreement exist with
the experimental findings summarized in subsection~\ref{sec:upair}.
The three percent weight of the $s$-type pairing inferred from experiments
in YBCO
\cite{K06}
might then be used as a constraint for tuning the values of the
phenomenological hopping energy parameters of the model.
\bigskip
\par
{\centering\section{CONCLUSIONS\label{sec:concl}}}
The scrutiny of the rigorous GMFA-GF solution of the Hamiltonian of the
two-dimensional two-band Hubbard model of the superconducting phase
transitions in cuprates
\cite{Pl95}
unveiled the occurrence of infinite quantities, stemming from hopping, in
the matrix elements of the energy matrix
$\tilde\mathcal{E}\sb{\sigma}({\bf q})$, Eq.~(\ref{eq:Esqdef}), in the limit
of the vanishing doping rates $\delta\rightarrow 0$.
\par
An analysis of the essential features of the cuprates which follow both from
experimental data and general theoretical representations shows that the
abovementioned infinities originate in the procedure of deriving the
Hamiltonian of the model, where the doping induced absence of part of the
electron orbitals in the CuO$\sb{2}$ plane was ignored.
A phenomenological approach to the inclusion of this feature into the model
resulted in the modification of the effective hopping contribution to the
Hamiltonian~(\ref{eq:H}), namely,
\begin{equation}
    H\sp{\prime}\sb{h} = \chi\sb{1}\chi\sb{2}H\sb{h}\, ,
\label{eq:Hhm}
\end{equation}
with $H\sb{h}$ given by~(\ref{eq:Hh}).
Technically, this result may be viewed as a renormalization of the Wannier
coefficients entering the Hubbard 1-forms~(\ref{eq:taup}) with the factor
$\chi\sb{1}\chi\sb{2}$ which expresses the availability of the spin states
for fermion hopping transitions inside the hopping conduction neighbourhood
of the reference spin lattice site~$i$.
\par
This leads, in the $({\bf r}, t)$-representation, to expressions of
the hopping terms of the energy matrix $\tilde\mathcal{E}\sb{ij\sigma}$
which simply multiply by a factor $\chi\sb{1}\chi\sb{2}$ the corresponding
quantities which have been obtained from $H\sb{h}$, Eq.~(\ref{eq:Hh}).
\par
The rigorous reduction of the order of correlation of the boson-boson
statistical averages involving singlet hopping (normal hopping and anomalous
charge-charge hopping correlations) brings supplementary
$\chi\sb{1}\chi\sb{2}$ factors in the resulting expressions.
\par
As a consequence, the normal correlation terms~(\ref{eq:Eqs}) as well as
the anomalous correlation terms~(\ref{eq:Fqs}) remain finite at any doping
rates $\delta$ both for the hole-doped and the electron-doped cuprates.
\par
The detailed investigation of the consequences of this modification of the
Hamiltonian of the two-dimensional two-band Hubbard model will be discussed
elsewhere.
\par
\medskip
{\centering\subsubsection*{Acknowledgments}}
\noindent
Partial financial support was secured by Romanian Authority for Scientific
Research (Project 7/2006 SIMFAP).
\bigskip
\begin{center}

\end{center}
\end{document}